# Exploring Paper as a Material:
# Plotting the Design Space of The Fabrication for Dynamic Paper-Based Interactions

**Ruhan Yang**

ATLAS Institute, University of Colorado Boulder

ruhan.yang@colorado.edu

**Ellen Yi-Luen Do**

ATLAS Institute, University of Colorado Boulder

ellen.do@colorado.edu

We reviewed 43 papers to understand the fabrication of dynamic paper-based interactions. We used a design space to classify tool selection, technique choice, and exploration of paper as a material. We classified 9 dimensions for the design space, including 4 dimensions for tools (precision, accommodation, complexity, and availability), 3 dimensions for techniques (cutting techniques, folding techniques, and integration techniques), and 2 dimensions for paper as the material (paper weight and paper type). The patterns we observed in the design space indicate a majority use of high precision tools, high complexity tools, and surface integration techniques in previous practice. Meanwhile, printing and plain paper are the leading material choices. We analyze these patterns and suggest potential directions for future work. Our study helps researchers locate different fabrication approaches and instances, thus fostering innovation in the field of paper-based interaction.

**Keywords**: Paper, paper-based interaction, paper computing, fabrication, design space.

# 1 INTRODUCTION

Since the early 1990s, as computing moved from screen-based to tangible interaction, researchers have investigated the use of paper as a medium. The unique advantages of paper make it an attractive medium for interaction: it is inexpensive, readily available, and familiar. This has led to numerous paper-based interaction projects. Early work in this realm focused on connecting the digital and physical worlds, represented by Wellner's Digital Desk [1991]. Wrensch's Programmable Hinge proposed the integration of paper crafts with electronic components, such as sensors and actuators, to transform them into tangible interfaces [Wrensch, 1998]. Since then, researchers have explored the use of handmade paper processes to incorporate electronics [Coelho, 2009], the use of paper pliability to create movable paper crafts [Qi, 2012], and the use of cutting and folding to create paper structures with haptic feedback [Chang, 2020]. These efforts expand the horizons of paper-based interaction. However, these studies have not been systematically reviewed, and it is challenging for researchers to have a comprehensive understanding of the current landscape of this field. As diverse techniques are proposed, we see the need to synthesize existing work. To this end, we investigate paper-based interactions from a fabrication perspective.

In this paper, we review the fabrication of paper-based interactions. We found more than 600 papers related to paper-based interactions in the ACM Digital Library and the Wiley Online Library databases, and screened 43 of them to form our corpus. We focus on the fabrication process in the literature to answer three questions: What tools were used? What manipulation techniques were used? What paper properties were investigated? Answering these questions will help researchers and practitioners better understand the fabrication of paper-based interactions.

We developed a design space for the fabrication of paper-based interactions through a review of our corpus. This design space consists of nine dimensions in three facets that address tools (tool precision, accommodation, complexity, and availability), techniques (cutting, folding, and integration techniques), and paper as a material (paper weight and paper type). These dimensions provide guidance to researchers in exploring existing paper-based interaction techniques, fostering their future refinement and advancement. Lastly, we analyze the distribution patterns in the design space and suggest potential directions for future work. We hope that this study will help researchers understand the overall spectrum of making paper-based interactions, thus promoting new explorations and innovations in the field.

# 2 BACKGROUND

In this section, we overview the background information within the domain of dynamic paper-based interaction. We clarify what our study includes and what is beyond the scope of the study.

## 2.1 Dynamic Paper-Based Interaction

Dynamic paper-based interaction involves enhancing traditional paper material with interactive elements such as electronic components. This turns paper into dynamic interfaces, engaging users as more than readers of static printed content. This shift from passive reading to active participation redefines the role of paper as a medium for interaction. These paper-based interactions extended the traditional functions of paper for recording text and drawings with augmented reality technology [Wellner 1991; Johnson, 1993]. Wrensch and Eisenberg proposed the Programmable Hinge and the HyperGami program [Wrensch, 1998; Eisenberg, 1998]. They explore the potential and possibilities of enhanced interactions through integrating computational elements into paper crafts. Since then, researchers have considered paper as an exploratory medium such as crafting circuits on paper [Buechley, 2009; Siegel, 2010], embedding electronic components within paper substrates [Coelho, 2009], fabricating paper-based circuit boards and sensors [Siegel, 2010; Russo, 2011], engineering interactive paper-based devices [Saul, 2010; Oh, 2018], and fostering interactive artistic creations on paper [Tsuji, 2011].

We focus on dynamic paper-based interaction, including research related to paper circuits, paper electronics, and moveable paper craft. These works regard paper as a material for exploration, integrating interactive elements onto or within paper. Previous research explored the technical aspects of paper circuits [Qi, 2010] and moveable paper craft [Zhu, 2013; Annett, 2015]. While these studies provide insights into the design of dynamic paper-based interactions, our work specifically explores fabrication. Our definition of dynamic paper-based interaction intentionally excludes works that merely treat paper as a medium for writing and printing to augment reading experiences. Such work focuses on seamlessly bridging the digital and physical worlds [Han, 2021], but does not explore the potential of paper as a material.

## 2.2 Why Fabrication?

A discussion of the fabrication process of paper-based interactions is necessary and will benefit researchers, especially those who are new to the field. We propose the fabrication-specific design space to gain a deeper understanding of the landscape. This will help researchers find instances under different fabrication methods, and thus choose an appropriate approach to achieve their desired design.

Often a design can be fabricated in different ways. For example, Figure 1 shows two different methods to create Kresling patterns. One method involves using two cylindrical objects to twist thin-walled cylindrical paper, allowing for a low complexity fabrication [Kresling, 2002]. This method provides a straightforward although imprecise means of achieving the desired paper structure. However, precision is sometimes essential, especially for paper-based interfaces. Then, researchers can use digital cutters to pre-cut and score the paper before folding it [Chen, 2020]. This method is more complex but offers high precision in the folding process, ensuring that the final outcome aligns perfectly with their design.

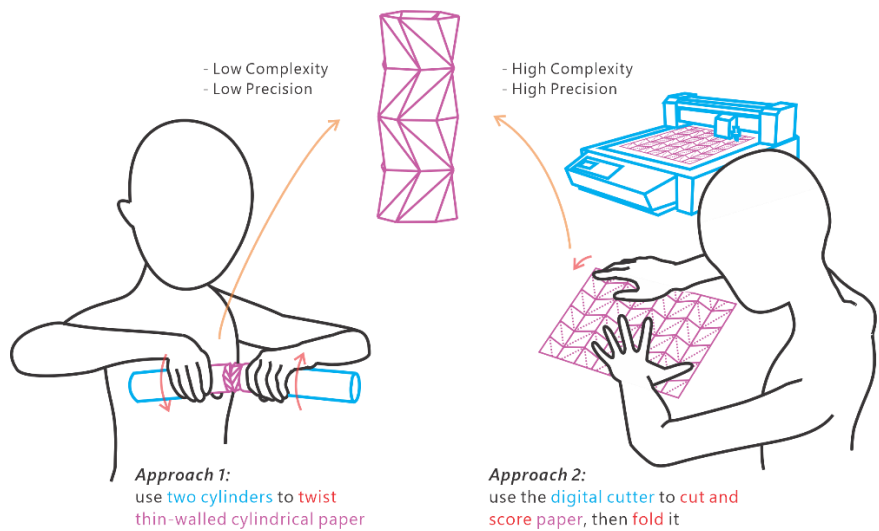

Figure 1: The Kresling pattern can be created using two different methods. Method 1 uses hand tools with low precision and low complexity, while Method 2 uses a digital cutter with high precision and higher complexity.

Existing paper-based interaction studies have suggested many different methods for adding conductive materials to paper, including painting directly on paper with conductive ink [Buechley, 2009], pasting conductive tape [Qi, 2010], using inkjet printers [Kawahara, 2013], using spray-coating tools [Thiemann, 2014], using screen printing [Brooke, 2023]. However, these fabrication approaches have not yet been reviewed and compiled. We systematically analyze three facets of fabrication, including the selection of tools, the choice of techniques, and the properties of paper as the material. Through our study, we organize and present different fabrication choices with examples. Our design space enables researchers and practitioners to learn about the spectrum of fabrication methods, thus helping them to make fabrication decisions.

The discussion of fabrication techniques is critical not only for implementation, but also for fostering innovation in the field. Through our study, researchers can gain an overview of existing methods, explore methods that have not been widely investigated, thus pushing the boundaries of what is possible with paper as an interactive medium.

## 3 METHODS

We used the PRISMA model [Liberati, 2009] to compile our corpus. The PRISMA (Preferred Reporting Items for Systematic Reviews and Meta-Analyses) model is a framework for conducting systematic reviews and meta-analyses. We began with a keyword-based search, followed by an investigation of related works by the authors our search revealed, while also incorporating exemplary cases. We identified, screened, and selected papers within the corpus by employing the four-phase flow diagram of the PRISMA model. Figure 2 below shows this process. We classified all papers (n=43) to identify key design space dimensions.

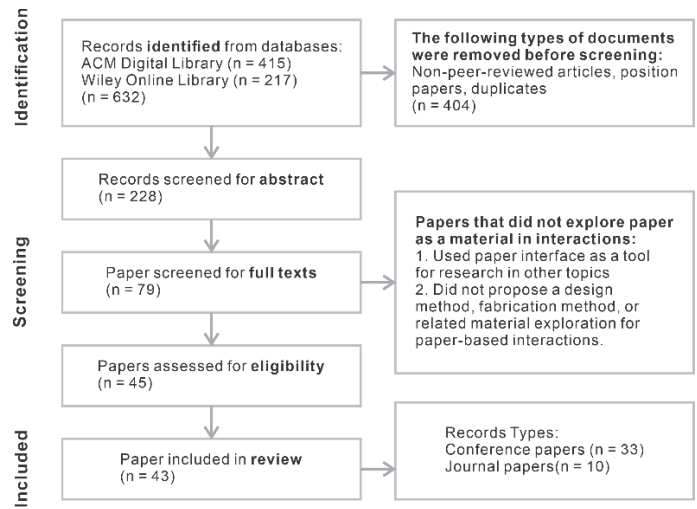

Figure 2: We used the PRISMA model [Liberati, 2009] to screen the literature.

### 3.1 Dataset and Criteria

We collected our dataset from ACM Digital Library and Wiley Online Library. ACM Digital Library primarily contains works closely related to human computer interaction, and Wiley Online Library features works more relevant to materials science. Both databases cover a range of tangible interaction design venues, including ACM conferences such as CHI, TEI, DIS, and the Advanced Materials series of journals. We also searched for relevant papers in the IEEE Xplore digital library; however, it did not yield content significantly aligned with our topic.

In order to collect papers relevant to paper-based interactions, we searched for papers where authors employed any of the following terms within their titles, abstracts, and keywords: "paper," "paper interaction," "paper interface," "paper computing," "paper electronic," "paper circuit," and "paper craft." We also searched for combinations of these keywords, such as "paper-based interaction" and "papercrafting".

Our initial quest yielded a total of 632 papers, with 415 sourced from the ACM Digital Library and 217 from the Wiley Online Library. Next, we removed duplicates, non-peer-reviewed articles, and position papers, which left us with a collection of 228 distinct works. During the third phase, we eliminated works that diverged from our main focus. We applied the following criteria: 1) the work must center on paper-based interactions and tangible paper interfaces; 2) The term "paper" must explicitly refer to physical sheets of paper; 3) The work must address the design, fabrication, or application of paper-based interactions. This led us to select 79 papers for subsequent full-text evaluation. We excluded 36 papers for two primary reasons: 1) The studies employed paper-based interactions as research tools without delving into the design and fabrication of paper-based interactions; 2) Novel design and fabrication methodologies for paper-based interactions were not proposed. This retrieval was concluded in August 2023, resulting in a corpus of 43 eligible works spanning from 1998 to 2023 sourced from ACM conferences (n=33), including CHI, TEI, UIST, DIS, and the Advanced Materials journal series (n=10).

### 3.2 Codebook

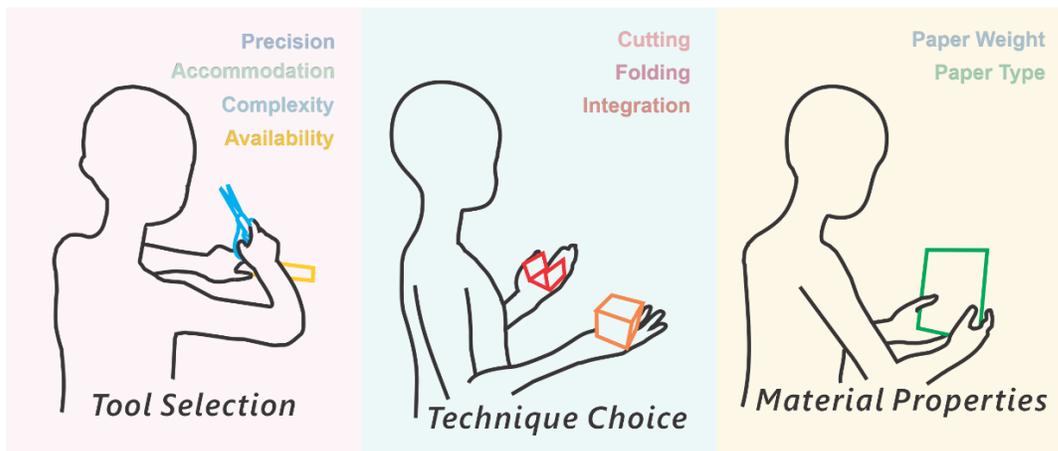

Figure 3: Our design space has three facets: Tool Selection (including precision, accommodation, complexity, and availability), Technique Choice (including cutting, folding, and integration technique), and Material Properties of paper (including paper weight and paper type.

We developed a codebook to classify the 43 papers, with nine dimensions to consider in three facets: tool selection, technique choice, and material properties. Figure 3 illustrates these three facets. Table 1 below shows our codebook. We provide a more detailed description with examples in Section 4 DESIGN SPACE.

Following this, we coded all the data within the corpus in order to classify the papers. The subcodes for each dimension are defined and discussed individually. We only analyze the actual fabrication process described in the paper and do not include the potential approaches they suggest. For example, in some papers [Annett, 2015 and Chang, 2020] that use the digital cutter but imply that a craft knife could also be used, we classified those as cases that use the digital cutter. Similarly, if a paper implies that different types of paper can be used [Li, 2019], we discuss only the kind of paper that has actually been tested and used. If a paper does not have enough information about fabrication, it may not be coded in some dimensions. If a paper provides more than one fabrication instance, it could have more than one code in the same dimension.

Table 1: Codebook

| Design Dimension | Codes and Description |
|---|---|
| **Tool Selection** | |
| Tool Precision | Low: Basic hand tools without great precision, e.g., scissors<br>Medium: Craft tools with better precision, e.g., craft knives<br>High: Digital controlled machines with good precision, e.g., laser cutters |
| Tool-Material Accommodation | Broad Accommodation: Tools can accommodate most paper thicknesses and types, e.g., screen printing<br>Limited Accommodation: Tools can accommodate a limited number of paper thicknesses and types, e.g., ink printers |
| Complexity | Low: Tools that are easy to use and require minimal skills<br>Moderate: Tools that require some experience<br>High: Tools that require advanced skills or expertise |
| Availability | High: Everyday tools that are commercially available<br>Medium: Household tools and machines<br>Low: Tools and machines not for household use or not commercially available |
| **Technique Choice** | |
| Cutting | Basic: Straight-line cutting and perforation<br>Advanced: Intricate pattern cutting |
| Folding | Basic: Folding with one or more non-overlapping straight lines<br>Advanced: Folding with multiple overlapping creases |
| Integration | Surface: Elements are applied to the surface of the paper<br>Embedded: Elements are embedded into the paper |
| **Material Properties** | |
| Paper Weight | Lightweight Paper: Paper that is less than 75 grams per square meter, or gsm<br>Printing Paper: Paper that is 75 gsm<br>Heavyweight Paper: Paper that is more than 75 gsm<br>Cardboard: Structure that has multiple layers of paper |
| Paper Type | Plain Paper: Paper that has not been specially treated and has a smooth surface and normal absorbency.<br>Special Surface: Papers with special surface treatments, including textured and coated papers<br>Special Absorbency: Paper with special absorbency, including high-absorbent cotton paper and low-absorbent cellulose paper |

In the following section 4 DESIGN SPACE, we introduce the design space in more detail. Within each dimension of the design space, we present examples from the relevant literature. In section 5 FINDINGS, we analyze distribution patterns in the design space, along with tables to help locate different fabrication instances.

## 4 DESIGN SPACE

From our coding, we identified 9 design dimensions describing the process of making paper-based interactions in three facets: tool selection, technique choice, and material properties. In this section, we provide an overview of this 9-dimensional design space. Figure 4 shows the distribution of those dimensions throughout our corpus.

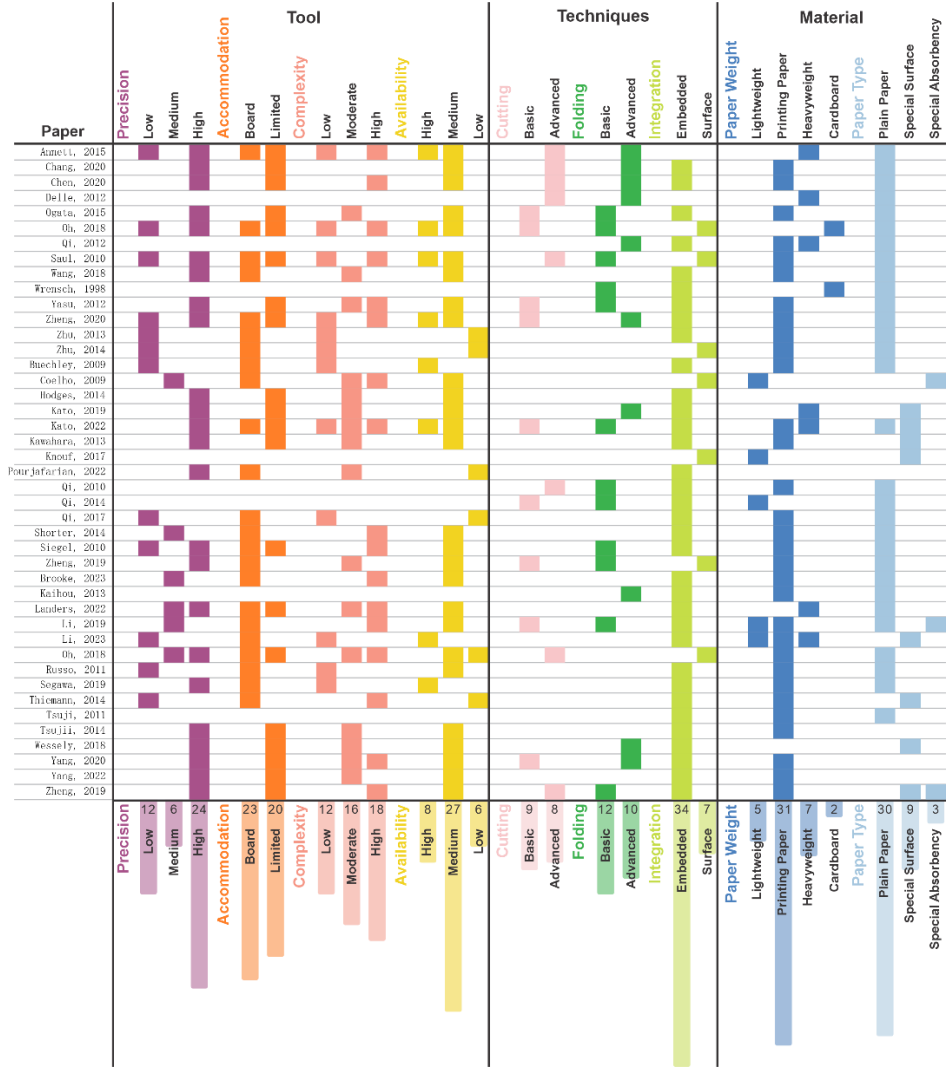

Figure 4: Our design space includes three facets: tools, techniques, and material properties. We map the corpus to this design space as shown in this diagram. The bar chart at the bottom shows the distribution of the design dimension.

### 4.1 Tool Selection

Tool selection talks about what was used to make paper-based interactions, focusing on the various characteristics of the tools used in fabrication. We established four design space dimensions for tools: Precision, Accommodation, Complexity, and Availability. Tables 2 to 6 show the detailed distribution for each dimension.

*4.1.1 Tool Precision*

Tool Precision refers to the level detail achievable during fabrication with a particular tool. We consider how well a tool can perform intricate operations on physical materials, evaluating whether a tool can consistently replicate intricate designs and patterns without damaging the materials. We classify three clusters of precision: low, medium, and high.

Table 2: Tools and their precision used in making paper-based interaction

| | Tool Type | Fabrication Instance |
|---|---|---|
| **Low Precision** (Basic hand tools) | Air Brush | Siegel, 2010 |
| | Code Stickers Toolkit (for paper circuit) | Qi, 2017 |
| | Craft Knife | Saul, 2010; Annett, 2015; Zheng, 2020 |
| | Paint Brush | Buechley, 2009 |
| | Pencil | Li, 2023 |
| | Rollerball Pen (with conductive silver ink) | Russo, 2011 |
| | Scissors | Oh, 2018; Zheng, 2020 |
| | Selective Induction Transmission Toolkit (for movable paper craft) | Zhu, 2013; Zhu, 2014 |
| | Spray-coat (with ZnO) | Thiemann, 2014 |
| **Medium Precision** (Craft tools) | Paper Making Tool | Coelho, 2009 |
| | Screen Printing Tool | Coelho, 2009; Shorter, 2014; Oh, 2018 Li, 2019; Landers, 2022; Brooke, 2023 |
| **High Precision** (Digital controlled machines) | Wax Printer | Landers, 2022 |
| | Desktop 3D Printer (Fused Deposition Modeling) | Zheng, 2019; Wang, 2018 |
| | Paper 3D Printer (Paper Sculpture Machine) | Oh, 2018 |
| | Laser Printer (with iron for transferring) | Segawa, 2019; Kato, 2022 |
| | Handheld Inkjet Printer | Pourjafarian, 2022 |
| | Digital Cutter | Saul, 2010; Siegel, 2010; Yasu, 2012; Annett, 2015; Oh, 2018; Zheng, 2019; Chang, 2020; Chen, 2020; Zheng, 2020; Yang, 2020; Kato, 2022; Landers, 2022 |
| | Inkjet Printer | Yasu, 2012; Kawahara, 2013; Tsujii, 2014; Hodges, 2014; Ogata, 2015; Wessely, 2018; Oh, 2018; Kato, 2019; Yang, 2020; Kato, 2022; Yang, 2022 |

Low precision tools are manual instruments without high precision, including scissors [Kato, 2022], adhesive tape [Qi, 2014], paint brushes [Buechley, 2009], roller pens [Russo, 2011], and pencils [Li, 2023].

Medium precision tools are craft tools with better precision, including paper making tools [Coelho, 2009], craft knives [Annett, 2015], and screen printing tools [Landers, 2022]. These tools are designed to be used for crafting thus enabling a higher level of precision than the basic tools, but they still do not achieve the precision that machines do.

High precision tools are digitally controlled machines with good precision. Numerous studies have investigated the use of desktop printers, including the inkjet printer [Kawahara, 2013] and the wax printer [Landers, 2022]. Other researchers also worked with diverse types of 3D printers [Wang, 2018 and Oh, 2018], as well as various cutting machines [Zheng, 2019 and Chang, 2020].

*4.1.2 Tool-Material Accommodation*

Tool-Material Accommodation examines how well the selected tool accommodates the material during the fabrication process. This dimension discusses whether the tools can perform tasks with various kinds of paper. We classify tool-material accommodation into two clusters: broad and limited. A broad accommodation indicates the tool is suitable for most paper thicknesses and types. Meanwhile, tools with limited accommodation are only suitable for a limited range of paper thicknesses and types.

Table 3: Tools and tool-material accommodation used in making paper-based interaction

|  | **Tool Type** | **Fabrication Instance** |
|---|---|---|
| **Broad Accommodation** (Capable of handling a wide range of paper) | Air Brush | Siegel, 2010 |
|  | Code Stickers Toolkit (for paper circuit) | Qi, 2017 |
|  | Craft Knife | Saul, 2010; Annett, 2015; Zheng, 2020 |
|  | Paint Brush | Buechley, 2009 |
|  | Pencil | Li, 2023 |
|  | Rollerball Pen (with conductive silver ink) | Russo, 2011 |
|  | Scissors | Oh, 2018; Zheng, 2020 |
|  | Selective Induction Transmission Toolkit (for movable paper craft) | Zhu, 2013; Zhu, 2014 |
|  | Spray-coat (with ZnO) | Thiemann, 2014 |
|  | Paper Making Tool | Coelho, 2009 |
|  | Screen Printing Tool | Coelho, 2009; Shorter, 2014; Oh, 2018 Li, 2019; Landers, 2022; Brooke, 2023 |
|  | Desktop 3D Printer (Fused Deposition Modeling) | Zheng, 2019; Wang, 2018 |
|  | Laser Printer (with iron for transferring) | Segawa, 2019; Kato, 2022 |
|  | Handheld Inkjet Printer | Pourjafarian, 2022 |

| | **Tool Type** | **Fabrication Instance** |
|---|---|---|
| **Limited Accommodation** (Capable of handling a limited range of paper) | Paper 3D Printer (Paper Sculpture Machine) | Oh, 2018 |
| | Wax Printer | Landers, 2022 |
| | Digital Cutter | Saul, 2010; Siegel, 2010; Yasu, 2012; Annett, 2015; Oh, 2018; Zheng, 2019; Chang, 2020; Chen, 2020; Zheng, 2020; Yang, 2020; Kato, 2022; Landers, 2022 |
| | Inkjet Printer | Yasu, 2012; Kawahara, 2013; Tsujii, 2014; Hodges, 2014; Ogata, 2015; Wessely, 2018; Oh, 2018; Kato, 2019; Yang, 2020; Kato, 2022; Yang, 2022; Pourjafarian, 2022 |

We've observed that most manual tools fall within the realm of broad accommodation, such as screen printing tools [Li, 2019]. However, for machines, particularly inkjet printers used for printing conductive materials, accommodation is more limited. The same goes for digital cutters. Such machines are usually not able to accommodate thicker paper and require adjustments based on sample fabrication outcomes [Brooke, 2023]. An intriguing instance involves the Mcor IRIS 3D printer in Oh's study [2018]. This device is engineered to perform 3D printing using US letter size or A4 office paper and is incompatible with other paper types and sizes. Meanwhile, two other cases that use 3D printers [Daudén, 2016 and Wang, 2018] are considered to be broadly accommodating as they can be used with a wider range of paper

*4.1.3 Tool Complexity*

Tool Complexity pertains to the level of expertise required to operate the selected tools. This dimension describes how user-friendly the tools are and how quickly individuals can become proficient in using them. We classify tool complexity into three clusters: low, moderate, and high.

Table 4: Tools and their complexity used in making paper-based interaction

| | **Tool Type** | **Fabrication Instance** |
|---|---|---|
| **Low Complexity** (require minimal skills) | Code Stickers Toolkit (for paper circuit) | Qi, 2017 |
| | Craft Knife | Saul, 2010; Annett, 2015; Zheng, 2020 |
| | Paint Brush | Buechley, 2009 |
| | Pencil | Li, 2023 |
| | Rollerball Pen (with conductive silver ink) | Russo, 2011 |
| | Scissors | Oh, 2018; Zheng, 2020 |
| | Selective Induction Transmission Toolkit (for movable paper craft) | Zhu, 2013; Zhu, 2014 |
| | Laser Printer (with iron for transferring) | Segawa, 2019; Kato, 2022 |

|  | **Tool Type** | **Fabrication Instance** |
|---|---|---|
| **Moderate Complexity** (require some experience) | Paper Making Tool | Coelho, 2009 |
|  | Desktop 3D Printer (Fused Deposition Modeling) | Zheng, 2019; Wang, 2018 |
|  | Wax Printer | Landers, 2022 |
|  | Handheld Inkjet Printer | Pourjafarian, 2022 |
|  | Inkjet Printer | Yasu, 2012; Kawahara, 2013; Tsujii, 2014; Hodges, 2014; Ogata, 2015; Wessely, 2018; Oh, 2018; Kato, 2019; Yang, 2020; Kato, 2022; Yang, 2022 |
| **High Complexity** (require advanced skills or expertise) | Air Brush | Siegel, 2010 |
|  | Spray-coat (with ZnO) | Thiemann, 2014 |
|  | Paper 3D Printer (Paper Sculpture Machine) | Oh, 2018 |
|  | Screen Printing Tool | Coelho, 2009; Shorter, 2014; Oh, 2018 Li, 2019; Landers, 2022; Brooke, 2023 |
|  | Digital Cutter | Saul, 2010; Siegel, 2010; Yasu, 2012; Annett, 2015; Oh, 2018; Zheng, 2019; Chang, 2020; Chen, 2020; Zheng, 2020; Yang, 2020; Kato, 2022; Landers, 2022 |

Tools of low complexity are easy to use and demand minimal skill, such as craft knives [Zheng, 2020] and iron [Segawa, 2019]. Toolkits [Zhu, 2013] that are designed for different paper-based interactions are also easy to use. Tools of moderate complexity require some prior experience or short training. Those tools include household machines, such as desktop printers [Tsujii, 2014] and regular 3D printers [Daudén, 2016]. Pourjafarian also proposed the use of handheld printers for crafting paper interfaces [2022]. High complexity tools require advanced skills. Those include professional manual tools like screen printing devices [Shorter, 2014], as well as digital machines like laser cutters and CNC machines [Saul, 2010]. Proficiency and expertise are essential for operating these tools.

*4.1.4 Tool Availability*

Tool Availability refers to the availability of the chosen tools. This dimension involves factors such as the cost, physical availability, and ease of procurement of the tools. We classify availability into three clusters: low, medium, and high.

High availability tools include basic hand tools such as scissors [Annett, 2015]. Tools of medium availability include household 3D printers [Daudén, 2016], digital cutters [Chen, 2020], specialized craft tools such as air brushes [Siegel, 2010], modified hand tools such as rollerball pens [Russo, 2011], and screen printing equipment [Yang, 2020] that use specialized commercially available printing materials. Tools with low availability include industrial-grade 3D printers [Oh, 2018], as well as toolkits [Qi, 2017] and printers [Pourjafarian, 2022] that are not commercially available.

Table 6: Tools and their availability used in making paper-based interaction

| | Tool Type | Fabrication Instance |
|---|---|---|
| **Low Availability** (Not for household use or not commercially available) | Code Stickers Toolkit (for paper circuit) | Qi, 2017 |
| | Selective Induction Transmission Toolkit (for movable paper craft) | Zhu, 2013; Zhu, 2014 |
| | Paper 3D Printer (Paper Sculpture Machine) | Oh, 2018 |
| | Handheld Inkjet Printer | Pourjafarian, 2022 |
| | Spray-coat (with ZnO) | Thiemann, 2014 |
| **Medium Availability** (Household tools and machines) | Air Brush | Siegel, 2010 |
| | Rollerball Pen (with conductive silver ink) | Russo, 2011 |
| | Paper Making Tool | Coelho, 2009 |
| | Wax Printer | Landers, 2022 |
| | Screen Printing Tool | Coelho, 2009; Shorter, 2014; Oh, 2018 Li, 2019; Landers, 2022; Brooke, 2023 |
| | Desktop 3D Printer (Fused Deposition Modeling) | Zheng, 2019; Wang, 2018 |
| | Digital Cutter | Saul, 2010; Siegel, 2010; Yasu, 2012; Annett, 2015; Oh, 2018; Zheng, 2019; Chang, 2020; Chen, 2020; Zheng, 2020; Yang, 2020; Kato, 2022; Landers, 2022 |
| | Inkjet Printer | Yasu, 2012; Kawahara, 2013; Tsujii, 2014; Hodges, 2014; Ogata, 2015; Wessely, 2018; Oh, 2018; Kato, 2019; Yang, 2020; Kato, 2022; Yang, 2022 |
| **High Availability** (Everyday tools) | Paint Brush | Buechley, 2009 |
| | Pencil | Li, 2023 |
| | Laser Printer (with iron for transferring) | Segawa, 2019; Kato, 2022 |
| | Scissors | Oh, 2018; Zheng, 2020 |
| | Craft Knife | Saul, 2010; Annett, 2015; Zheng, 2020 |

### 4.2 Technique Choice

Technique Choice talks about what to do with paper, focusing on the types of different manipulation techniques. We propose three dimensions based on the types of manipulations: Cutting, Folding, and Integration. Tables 7 to 9 show the distribution of our corpus in each design dimension.

*4.2.1 Cutting*

Cutting involves the removal of paper to shape it into the desired forms. The spectrum of cutting ranges from simple straight-line cuts to intricate patterns. We classify cutting techniques into two clusters: basic and advanced.

Table 7: Cutting techniques used in making paper-based interaction

|  | **Moveable paper crafts** | **Paper circuits** | **Paper electronics** |
|---|---|---|---|
| **Basic Cutting** (Straight-line cutting and perforation) | Yasu, 2012; Oh, 2018; Zheng, 2020; Ogata, 2015 | Zheng, 2019; Kato, 2022; Qi, 2014 | Li, 2019; Yang, 2020 |
| **Advanced Cutting** (Intricate pattern cutting) | Saul, 2010; Annett, 2015; Chang, 2020; Chen, 2020; Delle, 2012 | Qi, 2010 | Oh, 2018; Zheng, 2019 |

The basic technique includes straight-line cutting [Yasu, 2012] and perforation [Oh, 2018]. Both operations are easily performed using manual tools. Advanced cutting refers to intricate pattern cutting, including cuts involving multiple straight lines [Zheng, 2019] and curved lines [Chang, 2020], and dotted-line cutting [Chen, 2020]. It's more commonly observed in pop-up paper interfaces [Qi, 2010 and Delle, 2012].

*4.2.2 Folding*

Folding is another transformative process that alters the flatness of paper to create 3D structures. This dimension explores different ways of folding, from basic non-overlapping folding to more advanced approaches. The chosen folding technique influences the visual aesthetics, structural integrity, and the way users interact. We classify folding techniques into two clusters: basic and advanced.

Table 8: Folding techniques used in making paper-based interaction

|  | **Moveable paper crafts** | **Paper circuits** | **Paper electronics** |
|---|---|---|---|
| **Basic Folding** (With non-overlapping creases) | Wrensch, 1998; Saul, 2010; Yasu, 2012; Ogata, 2015; Oh, 2018 | Qi, 2010; Siegel, 2010; Qi, 2014; Zheng, 2019; Kato, 2022 | Zheng, 2019; Li, 2019 |
| **Advanced Folding** (With overlapping creases) | Delle, 2012; Qi, 2012; Annett, 2015; Chang, 2020; Chen, 2020; Zheng, 2020 | Kato, 2019 | Kaihou, 2013; Yang, 2020; Wessely, 2018 |

Basic folding techniques typically involve one or multiple straight-line creases, creating interfaces that can be opened and closed repeatedly [Wrensch, 1998 and Ogata, 2015], or connecting circuits in a three-dimensional manner [Siegel, 2010]. Advanced folding includes the folding of paper into interlocking structures [Wessely, 2018] and origami art pieces [Kaihou, 2013]. Functional paper structures, such as amplification mechanisms [Kato, 2019] and pressable structures [Chen, 2020 and Chang, 2020], also require advanced folding techniques.

*4.2.3 Integration*

Integration refers to combining paper with other materials or technologies to enhance functionality, involving embedding electronics, incorporating sensors, or using non-paper elements. Integration techniques determine the level of interactivity and the range of potential applications for the fabricated paper-based interactions. We classify integration into two clusters: surface and embedded integration

Table 9: Integration techniques used in paper-based interaction

|  | **Moveable paper crafts** | **Paper circuits** | **Paper electronics** |
|---|---|---|---|
| **Surface Integration** (Applied to the surface of the paper) | Zhu, 2013; Wang, 2018; Chang, 2020; Chen, 2020; Zheng, 2020; Yasu, 2012; Ogata, 2015; Qi, 2012; Wrensch, 1998 | Buechley, 2009; Qi, 2017; Shorter, 2014; Kato, 2019; Siegel, 2010; Kato, 2022; Hodges, 2014; Qi, 2010; Qi, 2014; Kawahara, 2013; Pourjafarian, 2022 | Li, 2019; Russo, 2011; Segawa, 2019; Li, 2023; Brooke, 2023; Yang, 2020; Zheng, 2019; Tsujii, 2014; Yang, 2022; Kaihou, 2013; Tsuji, 2011; Landers, 2022; Thiemann, 2014; Wessely, 2018; |
| **Embedded Integration** (Embedded into the paper) | Zhu, 2014; Saul, 2010; Oh, 2018 | Zheng, 2019; Coelho, 2009; Knouf, 2017 | Oh, 2018 |

Surface integration is the more common approach, involving the addition of conductive tape or coatings onto the paper surface. It has been used to create paper circuits [Qi, 2014], paper electronics [Thiemann, 2014], and paper batteries [Yang, 2022]. Wang also discusses incorporating 3D printing filaments onto the paper surface for creating paper actuators [2018]. Different elements, such as electronic components and shape memory alloys, can also be soldered or taped on the surface of paper [Qi, 2012].

Embedded integration includes the integration of external elements within the paper substrate, as well as the embedding of paper into other elements. Researchers create paper input devices by adding conductive material between layers of paper [Zhu, 2014 and Li, 2019]. Coelho first introduced the technique of embedding circuits within the paper during the hand papermaking process [2009], followed by Knouf's work on embedding circuits between layers of handmade paper [2017]. Oh proposed embedding circuits and electronic components within 3D paper structures [2018], while Daudén presented the embedding of circuits within 3D-printed structures [2016].

### 4.3 Material Properties

Material Properties talks about the physical attributes of the paper as a material, and how they impact the fabrication process and the resulting artifacts. We introduce two design dimensions related to the use of paper as a material: Weight and Type. Tables 10 and 11 show the distribution of paper features in those two dimensions. Some authors didn't specify their paper choices [Zheng, 2020]. For these situations, we used the accompanying illustrations from the papers to determine the material properties.

*4.3.1 Paper Weight*

Paper Weight refers to the thickness and density of the paper. Different weights affect how easily the paper can be manipulated. The thicker paper provides more structural integrity but may be challenging to fold intricately, while thinner paper allows for flexible designs but lacks durability. We classify paper weight into four clusters: lightweight paper, printing paper, heavyweight paper, and cardboard.

Table 10: The use of different weights of paper used in making paper-based interaction

|  | **Moveable paper crafts** | **Paper circuits** | **Paper electronics** |
|---|---|---|---|
| **Lightweight Paper** (Less than 75 gsm) |  | Coelho, 2009; Qi, 2014; Knouf, 2017 | Li, 2019; Li, 2023 |
| **Printing Paper** (75 gsm) | Saul, 2010; Yasu, 2012; Qi, 2012; Zhu, 2013; Zhu, 2014; Ogata, 2015; Wang, 2018; Chang, 2020; Chen, 2020; Zheng, 2020 | Buechley, 2009; Qi, 2010; Siegel, 2010; Qi, 2017; Zheng, 2019; Kato, 2022; Kawahara, 2013; Shorter, 2014 | Russo, 2011; Tsuji, 2011; Kaihou, 2013; Li, 2019; Thiemann, 2014; Tsujii, 2014; Oh, 2018; Segawa, 2019; Li, 2023; Yang, 2022; Brooke, 2023; Zheng, 2019 |
| **Heavyweight Paper** (More than 75 gsm) | Delle, 2012; Qi, 2012; Annett, 2015 | Kato, 2019 Kato, 2022 | Landers, 2022 Li, 2023 |
| **Cardboard** (Structure that has multiple layers of paper) | Wrensch, 1998; Oh, 2018 |  |  |

The most common paper type is printing paper (75 grams per square meter, or gsm), also known as office paper or copy paper. This type of paper is ubiquitous and easily manipulable, thus it is favored by many researchers [Siegel, 2010 and Kawahara, 2013]. The selection of lightweight paper (less than 75 gsm) options includes building on notebooks [Qi, 2014], as well as using handmade paper [Knouf, 2017]. Meanwhile, heavyweight paper (more than 75 gsm) is often represented by cardstock [Kato, 2022] and filter paper [Landers, 2022]. Cardboard has also been used to build moveable papercrafts [Oh, 2018]. Some interactions [Li 2019] could also be crafted using paper of any weight.

*4.3.2 Paper Type*

Paper Type focuses on distinctive characteristics between different types of paper, such as texture and absorbency. The selection of paper type affects how it integrates with other materials. We classify paper type into three clusters: plain paper, paper with special surface, and paper with special absorbency.

Table 11: The use of different types of paper used in making paper-based interaction

|  | **Moveable paper crafts** | **Paper circuits** | **Paper electronics** |
|---|---|---|---|
| **Plain Paper** (Has a smooth surface and normal absorbency) | Wrensch, 1998; Saul, 2010; Delle, 2012; Qi, 2012; Yasu, 2012; Qi, 2012; Zhu, 2013; Zhu, 2014; Annett, 2015; Ogata, 2015; Oh, 2018; Wang, 2018; Chang, 2020; Chen, 2020; Zheng, 2020 | Buechley, 2009; Qi, 2010 Siegel, 2010; Qi, 2014; Shorter, 2014; Qi, 2017; Zheng, 2019; Kato, 2022 | Russo, 2011; Tsuji, 2011; Kaihou, 2013; Oh, 2018; Li, 2019; Segawa, 2019; Landers, 2022; Brooke, 2023 |
| **Special Surface** |  | Knouf, 2017; Kato, 2019; Kato, 2022; Kawahara, 2013 | Tsujii, 2014; Zheng, 2019; Yang, 2020; Li, 2023; Thiemann, 2014 |
| **Special Absorbency** |  | Coelho, 2009 | Li, 2019; Yang, 2022 |

Plain paper is the most common choice, as most paper-based interactions were built with printing paper. Researchers have created paper interfaces using handmade paper [Knouf, 2017], which tends to have longer fibers and therefore a coarser texture. Some researchers have explored the use of coated papers, such as carbon-coated paper [Zheng, 2019], resin-coated paper [Kawahara, 2013], and ionogel-coated paper [Thiemann, 2014]. Other works [e.g., Tsujii, 2014] have also used sticker-coated paper. Additionally, researchers have utilized highly absorbent cotton paper [Coelho, 2009], Chinese rice paper [Li, 2019], and less absorbent cellulose paper [Yang, 2022] for creating paper circuits and paper electronics.

## 5 FINDINGS

In section 4 DESIGN SPACE, we discussed the distribution of the different fabrication factors in the 9 design dimensions. In this section, we discuss the distribution patterns we observed in the design space and provide insights into their possible causes.

### 5.1 Tool Selection

We observed that most of the tools being used are high precision. Table 12 shows the distribution of tool precision for making different types of interactions. There are 24 papers in our corpus that discuss the use of high precision tools for a total of 30 instances, a much higher number than the use of other tools (7 instances for medium precision and 13 instances for low precision). In terms of interaction types, 11 of these high precision tool instances are paper electronics, along with 9 paper circuits and 10 movable paper crafts. This suggests a strong preference for high precision in the current field of paper-based interaction.

The most common tool type is digital cutters (n=12). However, we also note possible alternatives for such high precision tools, such as the use of scissors and craft knives. Researchers have used scissors or craft knives in their practice [Annett, 2015 and Saul, 2010], or have suggested that users can use craft knives when they don't have access to a digital cutter [Chang, 2020]. Zheng [2020] stated that they used hand tools during their explorations, then used a digital cutter to get a higher presentation quality. Despite the high precision of digital cutters, we found it important to also consider their limitations, characterized by limited tool-material accommodation, high complexity, and medium availability. This implies that digital cutters may not be universally affordable to all individuals. In contrast, using hand tools could offer advantages in terms of higher accommodation, lower complexity, and higher availability.

Table 12: Tools and their precision used in making different types of paper-based interaction

|  | Moveable paper crafts | Paper circuits | Paper electronics |
|---|---|---|---|
| **Low Precision** (Basic hand tools) | **Craft Knife** Saul, 2010; Annett, 2015; Zheng, 2020 | **Air Brush** Siegel, 2010 | **Pencil** Li, 2023 |
|  | **Scissors** Oh, 2018; Zheng, 2020 | **Paint Brush** Buechley, 2009 | **Spray-coat (with ZnO)** Thiemann, 2014 |
|  | **Selective Induction Transmission Toolkit (for movable paper craft)** Zhu, 2013; Zhu, 2014 | **Code Stickers Toolkit (for paper circuit)** Qi, 2017 | **Rollerball Pen (with conductive silver ink)** Russo, 2011 |

|  | Moveable paper crafts | Paper circuits | Paper electronics |
|---|---|---|---|
| **Medium Precision** (Craft tools) |  | **Paper Making Tool** Coelho, 2009 **Screen Printing Tool** Coelho, 2009; Shorter, 2014 | **Screen Printing Tool** Oh, 2018 Li, 2019; Landers, 2022; Brooke, 2023 |
| **High Precision** (Digitally controlled machines) | **Desktop 3D Printer (Fused Deposition Modeling)** Wang, 2018 | **Desktop 3D Printer (Fused Deposition Modeling)** Zheng, 2019 | **Paper 3D Printer (Paper Sculpture Machine)** Oh, 2018 |
|  | **Inkjet Printer** Ogata, 2015; Yasu, 2012 | **Inkjet Printer** Kawahara, 2013; Hodges, 2014; Kato, 2019; Kato, 2022 | **Inkjet Printer** Oh, 2018; Yang, 2020; Wessely, 2018; Tsujii, 2014; Yang, 2022 |
|  | **Digital Cutter** Saul, 2010; Yasu, 2012; Annett, 2015; Oh, 2018; Chang, 2020; Chen, 2020; Zheng, 2020 | **Laser Printer (with iron for transferring)** Kato, 2022 | **Laser Printer (with iron for transferring)** Segawa, 2019 |
|  |  | **Digital Cutter** Siegel, 2010; Kato, 2022 | **Digital Cutter** Zheng, 2019; Yang, 2020; Landers, 2022 |
|  |  | **Handheld Inkjet Printer** Pourjafarian, 2022 | **Wax Printer** Landers, 2022 |

We encourage future researchers to explore a broader spectrum of tools beyond high precision. When dealing with intricate pattern cutting as demonstrated by Chen [2020], employing high precision tools can facilitate the creation of interfaces with speed and precision. However, for projects where precision is not as critical, diversifying the tool types in this domain could lead to more potential advantages. By exploring alternative tools like scissors and craft knives, individuals with limited resources can also engage in making paper-based interactions.

Another pattern we observed is that fewer cases involved the use of low complexity tools. Table 13 shows the distribution of tool complexity. We found a total of 13 papers in our corpus presented 14 instances of using low complexity tools, which is close to that of medium complexity tools (15 instances from 15 papers) and lower than that of high complexity tools (21 instances from 18 papers). We see this pattern is more obvious for paper electronics (3 out of 18 instances) and paper circuits (4 out of 15 instances). Inkjet printers of medium complexity are the most common tool choice. 11 of the cases involved the use of desktop printers, with a special case of Yasu proposing a handheld inkjet printer [2012]. Printer choices included the EPSON PX-S160T [Kato, 2019], the Epson ET-2550 [Wessely, 2018], the PIXUS iP100 [Tsujii, 2014], and the Cannon MG7530 [Yang, 2020]. Many researchers have used silver nanoparticle ink from Mitsubishi Paper Mills [Kato, 2019; Kato, 2022; Tsujii, 2014; Wessely, 2018; Yang, 2020], while Tsujii also used the QCR Solutions Corp thermo-chromic ink [2014].

In addition to inkjet printers, researchers have also explored the use of screen printing (n=6, high complexity) to print conductive ink onto paper. However, while inkjet printers have high precision, screen printing has only medium precision. Other tools, such as using a paint brush to paint [Buechley, 2009], have only low precision. We note that one potential alternative is gold foil transfer. Kato and Segawa used laser printers to print a pattern, and then transfer gold foil onto paper by ironing [Kato, 2022 and Segawa, 2019]. This method is more commonly used in craft communities, and is characterized by low complexity, high availability, and broad accommodation of different kinds of paper. Such cases imply that the use of low complexity tools does not necessarily lead to lower precision. We encourage researchers in the field to explore fabrication methods like gold foil transfer, which uses tools with low complexity and high availability to create paper interfaces. These methods enable fabrication to become approachable to a broader spectrum of individuals, thereby increasing the inclusivity and reach of paper-based interactions.

Table 13: Tools and their complexity used in making different types of paper-based interaction

|  | **Moveable paper crafts** | **Paper circuits** | **Paper electronics** |
|---|---|---|---|
| **Low Complexity** (Require minimal skills) | **Craft Knife** Saul, 2010; Annett, 2015; Zheng, 2020 | **Code Stickers Toolkit (for paper circuit)** Qi, 2017 | **Rollerball Pen (with conductive silver ink)** Russo, 2011 |
|  | **Scissors** Oh, 2018; Zheng, 2020 | **Paint Brush** Buechley, 2009 | **Pencil** Li, 2023 |
|  | **Selective Induction Transmission Toolkit (for movable paper craft)** Zhu, 2013; Zhu, 2014 | **Laser Printer (with iron for transferring)** Kato, 2022 | **Laser Printer (with iron for transferring)** Segawa, 2019 |
| **Moderate Complexity** (Require some experience) | **Inkjet Printer** Ogata, 2015; Yasu, 2012 | **Paper Making Tool** Coelho, 2009 | **Wax Printer** Landers, 2022 |
|  | **Desktop 3D Printer (Fused Deposition Modeling)** Wang, 2018 | **Handheld Inkjet Printer** Pourjafarian, 2022 | **Inkjet Printer** Oh, 2018; Yang, 2020; Wessely, 2018; Tsujii, 2014; Yang, 2022 |
|  |  | **Desktop 3D Printer (Fused Deposition Modeling)** Zheng, 2019 |  |
|  |  | **Inkjet Printer** Kawahara, 2013; Hodges, 2014; Kato, 2019; Kato, 2022 |  |
| **High Complexity** (Require advanced skills or expertise) | **Digital Cutter** Saul, 2010; Yasu, 2012; Annett, 2015; Oh, 2018; Chang, 2020; Chen, 2020; Zheng, 2020 | **Air Brush** Siegel, 2010 | **Spray-coat (with ZnO)** Thiemann, 2014 |
|  |  | **Screen Printing Tool** Coelho, 2009; Shorter, 2014 | **Paper 3D Printer (Paper Sculpture Machine)** Oh, 2018 |
|  |  | **Digital Cutter** Siegel, 2010; Kato, 2022 | **Screen Printing Tool** Oh, 2018; Li, 2019; Landers, 2022; Brooke, 2023 |
|  |  |  | **Digital Cutter** Zheng, 2019; Yang, 2020; Landers, 2022 |

## 5.2 Technique Choice

We found more projects that use surface integration techniques. Table 14 shows the distribution of surface and embedded integration for different interaction types, as well as the corresponding tool types. From our corpus, we found 34 works involving surface integration. We expect that this may be due to its alignment with the inherent characteristics and common uses of paper. People tend to think of paper as a "2D plane", and most of the time we only work on the surface of paper, such as writing and printing.

Table 14: Tools and integration techniques used in making paper-based interaction

|  | **Moveable paper crafts** | **Paper circuits** | **Paper electronics** |
|---|---|---|---|
| **Surface Integration** (Applied to the surface of the paper) | **Inkjet Printer** Ogata, 2015 | **Screen Printing Tool** Shorter, 2014 | **Wax Printer** Landers, 2022 |
|  | **Desktop 3D Printer (Fused Deposition Modeling)** Wang, 2018 | **Digital Cutter** Siegel, 2010 | **Spray-coat (with ZnO)** Thiemann, 2014 |
|  | **Digital Cutter** Yasu, 2012; Annett, 2015; Chang, 2020; Chen, 2020 | **Code Stickers Toolkit (for paper circuit)** Qi, 2017 | **Digital Cutter** Zheng, 2019 |
|  | **Selective Induction Transmission Toolkit (for movable paper craft)** Zhu, 2013 | **Handheld Inkjet Printer** Pourjafarian, 2022 | **Pencil** Li, 2023 |
|  |  | **Paint Brush** Buechley, 2009 | **Screen Printing Tool** Li, 2019; Brooke, 2023 |
|  |  | **Inkjet Printer** Kawahara, 2013; Hodges, 2014; Kato, 2019; Kato, 2022 | **Inkjet Printer** Yang, 2020; Wessely, 2018; Tsujii, 2014; Yang, 2022 |
|  |  |  | **Laser Printer (with iron for transferring)** Segawa, 2019 |
|  |  |  | **Rollerball Pen (with conductive silver ink)** Russo, 2011 |
| **Embedded Integration** (Embedded into the paper) | **Digital Cutter** Saul, 2010; Oh, 2018 | **Paper Making Tool** Coelho, 2009 | **Paper 3D Printer (Paper Sculpture Machine)** Oh, 2018 |
|  | **Selective Induction Transmission Toolkit (for movable paper craft)** Zhu, 2014 | **Desktop 3D Printer (Fused Deposition Modeling)** Zheng, 2019 |  |

However, we believe that embedded integration will open up more possibilities for paper-based interactions. We identified 7 efforts that used embedded integration spread across three interaction types, including 3 movable paper crafts [Saul, 2010; Zhu, 2014; Oh, 2018], 3 paper circuits [Coelho, 2009; Zheng, 2019; Knouf, 2017], and 1 paper electronics [Oh, 2018]. We classify these efforts into two forms. One form uses paper to build three-dimensional structures, as Saul added circuits to paper cubes [2010], Oh connected motors to cardboard and made movable mechanical structures [2018], and Zheng added folded paper circuits to 3D printed parts [2019]. The other form layers of paper together as a whole object, as when Zhu added coils between two pages of a book [2014], Coelho and Knouf added electronics between layers of paper [Coelho, 2009 and Knouf, 2017], and Oh used a special paper sculpture machine to cut and stack paper into three-dimensional objects [2018]. These embedded integrations add new forms of interaction to the paper-based interface, enriching the user experience. Thus, we call on future researchers to more thoroughly explore embedded integration techniques.

### 5.3 Material Properties

The prevalent selection of printing and plain paper represents the most significant pattern within the design space. In Table 15, we show the paper weight and type choices for different interaction types, and also list the types of tools involved in the fabrication. A total of 31 efforts used printing paper and paper of similar weight, and 30 efforts used plain paper. Paper-based interfaces emphasize the concept of ubiquity [Kawahara, 2013], so we believe that this distribution phenomenon is directly related to the ubiquity of printing and plain paper.

Table 15: The use of different kind of paper in efforts across different interaction types

|  | Moveable paper crafts | Paper circuits | Paper electronics |
|---|---|---|---|
| **Lightweight Paper** (Less than 75 gsm) |  | **Plain Paper** Qi, 2014 |  |
|  |  | **Special Surface** Knouf, 2017 | **Special Surface** Li, 2023 |
|  |  | **Special Absorbency** Coelho, 2009; | **Special Absorbency** Li, 2019 |
| **Printing Paper** (75 gsm) | **Plain Paper** Saul, 2010; Yasu, 2012; Qi, 2012; Zhu, 2013; Zhu, 2014; Ogata, 2015; Wang, 2018; Chang, 2020; Chen, 2020; Zheng, 2020 | **Plain Paper** Buechley, 2009; Siegel, 2010; Qi, 2010; Shorter, 2014; Zheng, 2019; Qi, 2017 | **Plain Paper** Russo, 2011; Tsuji, 2011; Kaihou, 2013; Oh, 2018; Segawa, 2019; Li, 2019; Brooke, 2023 |
|  |  | **Special Surface** Kawahara, 2013; Kato, 2022 | **Special Surface** Thiemann, 2014; Tsujii, 2014; Zheng, 2019; Yang, 2020; Li, 2023 |
|  |  |  | **Special Absorbency** Yang, 2022 |
| **Heavyweight Paper** (More than 75 gsm) | **Plain Paper** Delle, 2012; Qi, 2012; Annett, 2015 | **Plain Paper** Kato, 2022 | **Plain Paper** Landers, 2022 |
|  |  | **Special Absorbency** Kato, 2019; Li, 2023 |  |
| **Cardboard** (Structure that has multiple layers of paper) | **Plain Paper** Wrensch, 1998; Oh, 2018 |  |  |

The choice of cardboard and cardstock (heavyweight paper) is desired for paper interfaces that emphasize durability, such as movable paper crafts that perform back-and-forth movements [Wrensch, 1998; Qi, 2012; and Annett, 2015] and paper mechanical structures that are used in cooperation with other components [Delle, 2012 and Oh, 2018]. Lightweight paper is only used in paper electronics and paper circuits. Handmade paper in particular has longer fibers that can be easily processed and is the leading type of lightweight paper. Coelho explored adding electronics during the paper-making process [Coelho, 2009], Knouf added electronics between layers of hanji paper through the Joomchi process [Knouf, 2017], and Li's work used Chinese rice paper for its thin thickness and good printability [Li, 2019]. Coated paper is a common material in special paper types. Zheng created paper sensors by engraving and cutting carbon-coated paper [2019]. Meanwhile, resin-coated paper is favored for printing nanosilver ink with inkjet printers [Yang, 2020; Kato, 2022; Kato, 2019; Li, 2023; Kawahara, 2013]. This is a type of photographic paper with a surface coating of polyethylene or other resin. This coating prevents the ink from absorbing into the paper fibers, thus allowing the nanosilver ink to dry faster without spreading or requiring a long wait for it to cure. Cellulose paper has also been utilized in making paper electronics [Thiemann, 2014 and Yang, 2022]. This paper is absorbent and fluids are able to penetrate into the paper fibers. This allows cellulose paper base electronics to achieve a smaller bending radius and is more suitable for making flexible electronic devices. These examples demonstrate the potential of specialized paper types, which implies that there are more opportunities to explore unique kinds of paper for better fabrication results.

We believe that there are many more papers with special characteristics, and they could be utilized to their unique advantage in paper-based interactions. By constantly exploring and experimenting with various paper types, researchers could gain a deeper understanding of paper as a material. This will advance the field of paper-based interaction, expanding its design possibilities and extending more unique benefits.

## 6 DISCUSSION

### 6.1 Future Opportunities

Through our proposed design space, we break down each fabrication process into choices of tools, techniques, and materials. This helps us understand and communicate the various factors of making paper-based interactions. While we have briefly suggested potential research opportunities in our discussion of the design space, we summarize these here in three directions:

1, Explore tools with different precision, and experiment with more tools of low complexity. Paper-based interactions emphasize the affordances of paper as a material, encouraging creation and customization by users of different skill levels [Hodges, 2014; Annett, 2015; and Chang, 2020]. Choosing user-friendly tools is essential to foster inclusiveness within paper-based interactions. While maintaining the complexity of paper interactions, researchers can explore how to handle paper with tools that are low complexity and high availability. The use of gold foil transfers [Kato, 2022 and Segawa, 2019] are examples of how craft tools and techniques can be used for making paper-based interactions. This implies that more tools used in other fields could also be utilized for paper-based interfaces. Exploring different tools will lead to new design and fabrication possibilities in the field of paper-based interaction.

2, Explore more embedded integration approaches. Embedded integration allows for a more seamless and aesthetic integration of interactive elements in paper artifacts. Hiding interactive elements, such as sensors, LEDs, and conductive materials, within the structure will protect them from external factors, leading to more durable paper-based interactions. By treating paper as a three-dimensional object, researchers and designers could implement more complex and customized designs. Additionally, these hidden elements will be surprising to the user as these interactions are not visible on the paper surface, which offers unique and memorable user experiences.

3, Explore the material characteristics of paper. In the realm of paper-based interactions, researchers have primarily focused on limited kinds of paper as materials. Yet, we did not have a systematic exploration of paper as a material. This leads to a lack of insight into the material properties of different weights and types of paper. An interesting research direction could be to address the absorbency of paper, explore how it affects the performance of paper-based interactions, especially paper circuits and paper electronics. Meanwhile, the selection of paper type could also influence the visual and tactile quality of paper interfaces, thereby providing a unique sensory experience. In addition, we have noticed that some studies have defaulted to the term "paper" to refer to printing paper. In order to have a more in-depth investigation of the material properties of paper, we call on researchers to characterize the paper they use in future work.

## 6.2 Limitation

In this study, our focus is on the domain of dynamic paper-based interactions. Future research could potentially expand this scope to include areas such as AR-enhanced reading experiences, paper engineering, and more. We primarily collected our dataset by relying on keyword searches within the ACM Digital Library and Wiley Online Library, thus we may have missed relevant papers from other sources. Furthermore, our study focused only on published papers, while many paper-based interaction designs are practiced within enthusiast communities and may not take the form of academic papers. This bias could potentially impact the comprehensiveness of the design space. Further research should consider a more comprehensive approach that includes these paper-based interactions that have been shared across different communities.

Despite our efforts to ensure repeatability and reliability in our coding process, it's essential to acknowledge the subjectivity in its application. Our previous experience as researchers potentially influences our interpretation of the data. It's crucial to recognize that our clusters for some design dimensions may not universally apply across different communities. For instance, individuals within the crafts community, such as hobbyists, may perceive tools like 3D printers as more complex than screen printing tools. Therefore, we encourage readers to consider their own perspectives and expertise when using our design space of paper-based interactions.

## 7 CONCLUSION

We reviewed 43 papers to gain a deeper understanding of paper-based interactions, focusing on explorations of tools, techniques, and paper as a material. We developed a design space involving tool selection (tool precision, accommodation, complexity, and availability), technique choice (cutting, folding, and integration technique), and material properties of paper (paper weight and paper type). Researchers can use our design space to locate different fabrication approaches with related literature and instances.

Our survey identified several patterns in the current field of paper-based interaction. First, in terms of tool selection, most efforts use high precision tools and high complexity tools. In terms of technique choices, other interactive elements are often integrated onto the surface of the paper. In the exploration of paper as a material, most of the studies used printing paper and plain paper in their fabrication. We analyze these distribution patterns and suggest three possible directions for future work: 1) exploring tools with different precision and experimenting with low complexity tools; 2) exploring different embedded integration techniques; and 3) exploring the material properties of paper and experimenting with different kinds of paper. We hope this study could guide and inspire future developments in the field of paper-based interactions.

**REFERENCES**

Michelle Annett, Tovi Grossman, Daniel Wigdor, and George Fitzmaurice. 2015. MoveableMaker: Facilitating the Design, Generation, and Assembly of Moveable Papercraft. In Proceedings of the 28th Annual ACM Symposium on User Interface Software & Technology (UIST '15), 565–574. https://doi.org/10.1145/2807442.2807483

Robert Brooke, Jesper Edberg, Ioannis Petsagkourakis, Kathrin Freitag, Mohammad Yusuf Mulla, Marie Nilsson, Patrik Isacsson, and Peter Andersson Ersman. 2023. Paper Electronics Utilizing Screen Printing and Vapor Phase Polymerization. Advanced Sustainable Systems 7, 7: 2300058. https://doi.org/10.1002/adsu.202300058

Leah Buechley, Sue Hendrix, and Mike Eisenberg. 2009. Paints, paper, and programs: first steps toward the computational sketchbook. In Proceedings of the 3rd International Conference on Tangible and Embedded Interaction (TEI '09), 9–12. https://doi.org/10.1145/1517664.1517670

Zekun Chang, Tung D. Ta, Koya Narumi, Heeju Kim, Fuminori Okuya, Dongchi Li, Kunihiro Kato, Jie Qi, Yoshinobu Miyamoto, Kazuya Saito, and Yoshihiro Kawahara. 2020. Kirigami Haptic Swatches: Design Methods for Cut-and-Fold Haptic Feedback Mechanisms. In Proceedings of the 2020 CHI Conference on Human Factors in Computing Systems (CHI '20), 1–12. https://doi.org/10.1145/3313831.3376655

Christopher Chen, David Howard, Steven L. Zhang, Youngwook Do, Sienna Sun, Tingyu Cheng, Zhong Lin Wang, Gregory D. Abowd, and HyunJoo Oh. 2020. SPIN (Self-powered Paper Interfaces): Bridging Triboelectric Nanogenerator with Folding Paper Creases. In Proceedings of the Fourteenth International Conference on Tangible, Embedded, and Embodied Interaction (TEI '20), 431–442. https://doi.org/10.1145/3374920.3374946

Marcelo Coelho, Lyndl Hall, Joanna Berzowska, and Pattie Maes. 2009. Pulp-based computing: a framework for building computers out of paper. In CHI '09 Extended Abstracts on Human Factors in Computing Systems (CHI EA '09), 3527–3528. https://doi.org/10.1145/1520340.1520525

Claudia Daudén Roquet, Jeeeun Kim, and Tom Yeh. 2016. 3D Folded PrintGami: Transforming Passive 3D Printed Objects to Interactive by Inserted Paper Origami Circuits. In Proceedings of the 2016 ACM Conference on Designing Interactive Systems (DIS '16), 187–191. https://doi.org/10.1145/2901790.2901891

Stefano Delle Monache, Davide Rocchesso, Jie Qi, Leah Buechley, Amalia De Götzen, and Dario Cestaro. 2012. Paper mechanisms for sonic interaction. In Proceedings of the Sixth International Conference on Tangible, Embedded and Embodied Interaction (TEI '12), 61–68. https://doi.org/10.1145/2148131.2148146

Mike Eisenberg and Ann Nishioka Eisenberg. 1998. Shop Class for the Next Millenium: Education through Computer-Enriched Handicrafts. 1998, 2: Art. 8. https://doi.org/10.5334/1998-8

Feng Han, Yifei Cheng, Megan Strachan, and Xiaojuan Ma. 2021. Hybrid Paper-Digital Interfaces: A Systematic Literature Review. In Proceedings of the 2021 ACM Designing Interactive Systems Conference (DIS '21), 1087–1100. https://doi.org/10.1145/3461778.3462059

Steve Hodges, Nicolas Villar, Nicholas Chen, Tushar Chugh, Jie Qi, Diana Nowacka, and Yoshihiro Kawahara. 2014. Circuit stickers: peel-and-stick construction of interactive electronic prototypes. In Proceedings of the SIGCHI Conference on Human Factors in Computing Systems (CHI '14), 1743–1746. https://doi.org/10.1145/2556288.2557150

Walter Johnson, Herbert Jellinek, Leigh Klotz, Ramana Rao, and Stuart K. Card. 1993. Bridging the paper and electronic worlds: the paper user interface. In Proceedings of the INTERACT '93 and CHI '93 Conference on Human Factors in Computing Systems (CHI '93), 507–512. https://doi.org/10.1145/169059.169445

Tatsuya Kaihou and Akira Wakita. 2013. Electronic origami with the color-changing function. In Proceedings of the second international workshop on Smart material interfaces: another step to a material future (SMI '13), 7–12. https://doi.org/10.1145/2534688.2534690

Kunihiro Kato, Kaori Ikematsu, Yuki Igarashi, and Yoshihiro Kawahara. 2022. Paper-Woven Circuits: Fabrication Approach for Papercraft-based Electronic Devices. In Sixteenth International Conference on Tangible, Embedded, and Embodied Interaction (TEI '22), 1–11. https://doi.org/10.1145/3490149.3502253

Kunihiro Kato, Kazuya Saito, and Yoshihiro Kawahara. 2019. OrigamiSpeaker: Handcrafted Paper Speaker with Silver Nano-Particle Ink. In Extended Abstracts of the 2019 CHI Conference on Human Factors in Computing Systems (CHI EA '19), 1–6. https://doi.org/10.1145/3290607.3312872

Yoshihiro Kawahara, Steve Hodges, Benjamin S. Cook, Cheng Zhang, and Gregory D. Abowd. 2013. Instant inkjet

circuits: lab-based inkjet printing to support rapid prototyping of UbiComp devices. In Proceedings of the 2013 ACM international joint conference on Pervasive and ubiquitous computing (UbiComp '13), 363–372. https://doi.org/10.1145/2493432.2493486

Nicholas A. Knouf. 2017. Felted Paper Circuits Using Joomchi. In Proceedings of the Eleventh International Conference on Tangible, Embedded, and Embodied Interaction (TEI '17), 443–450. https://doi.org/10.1145/3024969.3025071

Biruta Kresling. 2002. B. Kresling (2002) Folded Tubes as Compared to Kikko ("Tortoise-Shell") Bamboo.

Mya Landers, Anwar Elhadad, Maryam Rezaie, and Seokheun Choi. 2022. Integrated Papertronic Techniques: Highly Customizable Resistor, Supercapacitor, and Transistor Circuitry on a Single Sheet of Paper. ACS Applied Materials & Interfaces 14, 40: 45658–45668. https://doi.org/10.1021/acsami.2c13503

Sen Li, Ning Pan, Zijie Zhu, Ruya Li, Baoqing Li, Jiaru Chu, Guanglin Li, Yu Chang, and Tingrui Pan. 2019. All-in-One Iontronic Sensing Paper. Advanced Functional Materials 29, 11: 1807343. https://doi.org/10.1002/adfm.201807343

Yiyang Li, Tianze Guo, Jiajun He, Jiazheng Yu, Hao Dong, Ting Zhang, and Guanyun Wang. 2023. Pencil-E: Crafting Functional Electronics Using Pencils and Paper. In Extended Abstracts of the 2023 CHI Conference on Human Factors in Computing Systems (CHI EA '23), 1–7. https://doi.org/10.1145/3544549.3585759

Masa Ogata and Masaaki Fukumoto. 2015. FluxPaper: Reinventing Paper with Dynamic Actuation Powered by Magnetic Flux. In Proceedings of the 33rd Annual ACM Conference on Human Factors in Computing Systems (CHI '15), 29–38. https://doi.org/10.1145/2702123.2702516

Hyunjoo Oh, Sherry Hsi, Michael Eisenberg, and Mark D. Gross. 2018. Paper mechatronics: present and future. In Proceedings of the 17th ACM Conference on Interaction Design and Children (IDC '18), 389–395. https://doi.org/10.1145/3202185.3202761

Hyunjoo Oh, Tung D. Ta, Ryo Suzuki, Mark D. Gross, Yoshihiro Kawahara, and Lining Yao. 2018. PEP (3D Printed Electronic Papercrafts): An Integrated Approach for 3D Sculpting Paper-Based Electronic Devices. In Proceedings of the 2018 CHI Conference on Human Factors in Computing Systems (CHI '18), 1–12. https://doi.org/10.1145/3173574.3174015

Matthew J Page, Joanne E McKenzie, Patrick M Bossuyt, Isabelle Boutron, Tammy C Hoffmann, Cynthia D Mulrow, Larissa Shamseer, Jennifer M Tetzlaff, Elie A Akl, Sue E Brennan, Roger Chou, Julie Glanville, Jeremy M Grimshaw, Asbjørn Hróbjartsson, Manoj M Lalu, Tianjing Li, Elizabeth W Loder, Evan Mayo-Wilson, Steve McDonald, Luke A McGuinness, Lesley A Stewart, James Thomas, Andrea C Tricco, Vivian A Welch, Penny Whiting, and David Moher. 2021. The PRISMA 2020 statement: an updated guideline for reporting systematic reviews. BMJ: n71. https://doi.org/10.1136/bmj.n71

Narjes Pourjafarian, Marion Koelle, Fjolla Mjaku, Paul Strohmeier, and Jürgen Steimle. 2022. Print-A-Sketch: A Handheld Printer for Physical Sketching of Circuits and Sensors on Everyday Surfaces. In CHI Conference on Human Factors in Computing Systems (CHI '22), 1–17. https://doi.org/10.1145/3491102.3502074

Jie Qi and Leah Buechley. 2010. Electronic popables: exploring paper-based computing through an interactive pop-up book. In Proceedings of the fourth international conference on Tangible, embedded, and embodied interaction (TEI '10), 121–128. https://doi.org/10.1145/1709886.1709909

Jie Qi and Leah Buechley. 2012. Animating paper using shape memory alloys. In Proceedings of the SIGCHI Conference on Human Factors in Computing Systems (CHI '12), 749–752. https://doi.org/10.1145/2207676.2207783

Jie Qi and Leah Buechley. 2014. Sketching in circuits: designing and building electronics on paper. In Proceedings of the SIGCHI Conference on Human Factors in Computing Systems (CHI '14), 1713–1722. https://doi.org/10.1145/2556288.2557391

Jie Qi, Asli Demir, and Joseph A. Paradiso. 2017. Code Collage: Tangible Programming On Paper With Circuit Stickers. In Proceedings of the 2017 CHI Conference Extended Abstracts on Human Factors in Computing Systems (CHI EA '17), 1970–1977. https://doi.org/10.1145/3027063.3053084

Analisa Russo, Bok Yeop Ahn, Jacob J. Adams, Eric B. Duoss, Jennifer T. Bernhard, and Jennifer A. Lewis. 2011. Pen-on-Paper Flexible Electronics. Advanced Materials 23, 30: 3426–3430. https://doi.org/10.1002/adma.201101328

Greg Saul, Cheng Xu, and Mark D. Gross. 2010. Interactive paper devices: end-user design & fabrication. In Proceedings of the fourth international conference on Tangible, embedded, and embodied interaction (TEI '10), 205–212. https://doi.org/10.1145/1709886.1709924

Norihisa Segawa, Kunihiro Kato, and Hiroyuki Manabe. 2019. Rapid Prototyping of Paper Electronics Using a Metal Leaf and Laser Printer. In The Adjunct Publication of the 32nd Annual ACM Symposium on User Interface Software and Technology (UIST '19), 99–101. https://doi.org/10.1145/3332167.3356885

Michael Shorter, Jon Rogers, and John McGhee. 2014. Enhancing everyday paper interactions with paper circuits. In Proceedings of the 2014 conference on Designing interactive systems (DIS '14), 39–42. https://doi.org/10.1145/2598510.2598584

Adam C. Siegel, Scott T. Phillips, Michael D. Dickey, Nanshu Lu, Zhigang Suo, and George M. Whitesides. 2010. Foldable Printed Circuit Boards on Paper Substrates. Advanced Functional Materials 20, 1: 28–35. https://doi.org/10.1002/adfm.200901363

Stefan Thiemann, Swetlana J. Sachnov, Fredrik Pettersson, Roger Bollström, Ronald Österbacka, Peter Wasserscheid, and Jana Zaumseil. 2014. Cellulose-Based Ionogels for Paper Electronics. Advanced Functional Materials 24, 5: 625–634. https://doi.org/10.1002/adfm.201302026

Kohei Tsuji and Akira Wakita. 2011. Anabiosis: an interactive pictorial art based on polychrome paper computing. In Proceedings of the 8th International Conference on Advances in Computer Entertainment Technology - ACE '11, 1. https://doi.org/10.1145/2071423.2071521

Takahiro Tsujii, Naoya Koizumi, and Takeshi Naemura. 2014. Inkantatory paper: dynamically color-changing prints with multiple functional inks. In Proceedings of the adjunct publication of the 27th annual ACM symposium on User interface software and technology (UIST'14 Adjunct), 39–40. https://doi.org/10.1145/2658779.2659103

Guanyun Wang, Tingyu Cheng, Youngwook Do, Humphrey Yang, Ye Tao, Jianzhe Gu, Byoungkwon An, and Lining Yao. 2018. Printed Paper Actuator: A Low-cost Reversible Actuation and Sensing Method for Shape Changing Interfaces. In Proceedings of the 2018 CHI Conference on Human Factors in Computing Systems, 1–12. https://doi.org/10.1145/3173574.3174143

Pierre Wellner. 1991. The DigitalDesk calculator: tangible manipulation on a desk top display. In Proceedings of the 4th annual ACM symposium on User interface software and technology (UIST '91), 27–33. https://doi.org/10.1145/120782.120785

Michael Wessely, Nadiya Morenko, Jürgen Steimle, and Michael Schmitz. 2018. Interactive Tangrami: Rapid Prototyping with Modular Paper-folded Electronics. In The 31st Annual ACM Symposium on User Interface Software and Technology Adjunct Proceedings (UIST '18 Adjunct), 143–145. https://doi.org/10.1145/3266037.3271630

Thomas Wrensch and Michael Eisenberg. 1998. The programmable hinge: toward computationally enhanced crafts. In Proceedings of the 11th annual ACM symposium on User interface software and technology (UIST '98), 89–96. https://doi.org/10.1145/288392.288577

Peihua Yang, Jia Li, Seok Woo Lee, and Hong Jin Fan. 2022. Printed Zinc Paper Batteries. Advanced Science 9, 2: 2103894. https://doi.org/10.1002/advs.202103894

Tilo H. Yang, Hikaru Hida, Daiki Ichige, Jun Mizuno, C. Robert Kao, and Jun Shintake. 2020. Foldable Kirigami Paper Electronics. physica status solidi (a) 217, 9: 1900891. https://doi.org/10.1002/pssa.201900891

Kentaro Yasu and Masahiko Inami. 2012. POPAPY: Instant Paper Craft Made Up in a Microwave Oven. In Advances in Computer Entertainment, Anton Nijholt, Teresa Romão and Dennis Reidsma (eds.). Springer Berlin Heidelberg, Berlin, Heidelberg, 406–420. https://doi.org/10.1007/978-3-642-34292-9_29

Clement Zheng, Peter Gyory, and Ellen Yi-Luen Do. 2020. Tangible Interfaces with Printed Paper Markers. In Proceedings of the 2020 ACM Designing Interactive Systems Conference. Association for Computing Machinery, New York, NY, USA, 909–923. Retrieved February 11, 2022 from https://doi.org/10.1145/3357236.3395578

Clement Zheng, HyunJoo Oh, Laura Devendorf, and Ellen Yi-Luen Do. 2019. Sensing Kirigami. In Proceedings of the 2019 on Designing Interactive Systems Conference (DIS '19), 921–934. https://doi.org/10.1145/3322276.3323689

Kening Zhu. 2014. PaperIO: Paper-Based 3D I/O Interface Using Selective Inductive Power Transmission. In Distributed, Ambient, and Pervasive Interactions, Norbert Streitz and Panos Markopoulos (eds.). Springer International Publishing, Cham, 217–228. https://doi.org/10.1007/978-3-319-07788-8_21

Kening Zhu and Shengdong Zhao. 2013. AutoGami: a low-cost rapid prototyping toolkit for automated movable paper craft. In Proceedings of the SIGCHI Conference on Human Factors in Computing Systems (CHI '13), 661–670. https://doi.org/10.1145/2470654.2470748